\documentclass[10pt,conference,twoside, a4paper]{IEEEtran}

\usepackage{cite}
\usepackage[T1]{fontenc}
\usepackage{graphicx}
\usepackage{amssymb}
\usepackage{amsmath}
\usepackage{paralist}
\usepackage{setspace}
\usepackage{microtype}
\usepackage{hyperref}
\usepackage{url}
\usepackage{balance}
\usepackage[caption=false,font=footnotesize]{subfig}
\usepackage{xcolor}
\usepackage{accents}
\usepackage{dsfont}
\usepackage{lipsum}
\usepackage{booktabs}
\usepackage[nolist,nohyperlinks]{acronym}

\usepackage{framed}
\usepackage{nicefrac}
\usepackage{multirow}
\usepackage{algorithm}
\usepackage{algpseudocode}

\newtheorem{alg}{Algorithm}

\acrodef{AMAM}[AM/AM]{amplitude modulation/amplitude modulation}
\acrodef{AMPM}[AM/PM]{amplitude modulation/phase modulation}
\acrodef{MU}[MU]{multi-user}
\acrodef{MIMO}[MIMO]{multiple-input multiple-output}
\acrodef{ZF}[ZF]{zero forcing}
\acrodef{MRT}[MRT]{maximum-ratio transmission}
\acrodef{DFT}[DFT]{discrete Fourier transform}
\acrodef{IDFT}[IDFT]{inverse discrete Fourier transform}
\acrodef{RF}[RF]{radio frequency}
\acrodef{PA}[PA]{power amplifier}
\acrodef{PAR}[PAR]{peak-to-average power ratio}
\acrodef{NR}[NR]{new radio}
\acrodef{LTE}[LTE]{long-term evolution}
\acrodef{LO}[LO]{local oscillator}
\acrodef{UE}[UE]{user equipment}
\acrodef{BS}[BS]{base station}
\acrodef{SNR}[SNR]{signal-to-noise ratio}
\acrodef{SINDR}[SINDR]{signal-to-interference-noise-and-distortion ratio}
\acrodef{PLL}[PLL]{phased-locked Loop}
\acrodef{GMP}[GMP]{generalized memory polynomial}
\acrodef{ADC}[ADC]{analog-to-digital converter}
\acrodef{DAC}[DAC]{digital-to-analog converter}
\acrodef{TDD}[TDD]{time-division duplex}
\acrodef{AWGN}[AWGN]{additive white Gaussian noise}
\acrodef{WSS}[WSS]{wide-sense stationary}
\acrodef{LNA}[LNA]{low-noise amplifier}
\acrodef{CSI}[CSI]{channel state information}
\acrodef{OFDM}[OFDM]{orthogonal frequency-division multiplexing}
\acrodef{OSR}[OSR]{oversampling ratio}
\acrodef{LMMSE}[LMMSE]{linear minimum mean square error}
\acrodef{RHS}[RHS]{right hand side}
\acrodef{PSD}[PSD]{power spectral density}
\acrodef{OOB}[OOB]{out of band}
\acrodef{ICI}[ICI]{intercarrier interference}
\acrodef{FFT}[FFT]{Fast Fourier Transform}
\acrodef{IFFT}[IFFT]{Inverse Fast Fourier Transform}
\acrodef{LSB}[LSB]{least significant bit}
\acrodef{ENOB}[ENOB]{effective number of bits}
\acrodef{MSE}[MSE]{mean square error}
\acrodef{BER}[BER]{bit-error rate}
\acrodef{QPSK}[QPSK]{quadrature phase-shift keying}
\acrodef{QAM}[QAM]{quadrature amplitude modulation}
\acrodef{WF}[WF]{Wiener-filter}
\acrodef{WFQ}[WFQ]{WF-quantized}
\acrodef{SQUID}[SQUID]{squared-infinity-norm Douglas-Rachford splitting}
\acrodef{EVM}[EVM]{error-vector magnitude}
\acrodef{CDF}[CDF]{cumulative distribution function}
\acrodef{CCDF}[CCDF]{complementary cumulative distribution function}

\newcommand{\snr}{\rho}

\usepackage{amssymb}
\usepackage{amsfonts}
\usepackage{mathrsfs}
\usepackage{xspace}
\usepackage{bm}
\usepackage{upgreek}

\newcommand{\safemath}[2]{\newcommand{#1}{\ensuremath{#2}\xspace}}

\safemath{\bma}{\mathbf{a}}
\safemath{\bmb}{\mathbf{b}}
\safemath{\bmc}{\mathbf{c}}
\safemath{\bmd}{\mathbf{d}}
\safemath{\bme}{\mathbf{e}}
\safemath{\bmf}{\mathbf{f}}
\safemath{\bmg}{\mathbf{g}}
\safemath{\bmh}{\mathbf{h}}
\safemath{\bmi}{\mathbf{i}}
\safemath{\bmj}{\mathbf{j}}
\safemath{\bmk}{\mathbf{k}}
\safemath{\bml}{\mathbf{l}}
\safemath{\bmm}{\mathbf{m}}
\safemath{\bmn}{\mathbf{n}}
\safemath{\bmo}{\mathbf{o}}
\safemath{\bmp}{\mathbf{p}}
\safemath{\bmq}{\mathbf{q}}
\safemath{\bmr}{\mathbf{r}}
\safemath{\bms}{\mathbf{s}}
\safemath{\bmt}{\mathbf{t}}
\safemath{\bmu}{\mathbf{u}}
\safemath{\bmv}{\mathbf{v}}
\safemath{\bmw}{\mathbf{w}}
\safemath{\bmx}{\mathbf{x}}
\safemath{\bmy}{\mathbf{y}}
\safemath{\bmz}{\mathbf{z}}
\safemath{\bmzero}{\mathbf{0}}
\safemath{\bmone}{\mathbf{1}}

\bmdefine{\biad}{a}
\bmdefine{\bibd}{b}
\bmdefine{\bicd}{c}
\bmdefine{\bidd}{d}
\bmdefine{\bied}{e}
\bmdefine{\bifd}{f}
\bmdefine{\bigd}{g}
\bmdefine{\bihd}{h}
\bmdefine{\biid}{i}
\bmdefine{\bijd}{j}
\bmdefine{\bikd}{k}
\bmdefine{\bild}{l}
\bmdefine{\bimd}{m}
\bmdefine{\bind}{n}
\bmdefine{\biod}{o}
\bmdefine{\bipd}{p}
\bmdefine{\biqd}{q}
\bmdefine{\bird}{r}
\bmdefine{\bisd}{s}
\bmdefine{\bitd}{t}
\bmdefine{\biud}{u}
\bmdefine{\bivd}{v}
\bmdefine{\biwd}{w}
\bmdefine{\bixd}{x}
\bmdefine{\biyd}{y}
\bmdefine{\bizd}{z}

\bmdefine{\bixid}{\xi}
\bmdefine{\bilambdad}{\lambda}
\bmdefine{\bimud}{\mu}
\bmdefine{\bithetad}{\theta}
\bmdefine{\biphid}{\phi}
\bmdefine{\bideltad}{\delta}

\safemath{\bmia}{\biad}
\safemath{\bmib}{\bibd}
\safemath{\bmic}{\bicd}
\safemath{\bmid}{\bidd}
\safemath{\bmie}{\bied}
\safemath{\bmif}{\bifd}
\safemath{\bmig}{\bigd}
\safemath{\bmih}{\bihd}
\safemath{\bmii}{\biid}
\safemath{\bmij}{\bijd}
\safemath{\bmik}{\bikd}
\safemath{\bmil}{\bild}
\safemath{\bmim}{\bimd}
\safemath{\bmin}{\bind}
\safemath{\bmio}{\biod}
\safemath{\bmip}{\bipd}
\safemath{\bmiq}{\biqd}
\safemath{\bmir}{\bird}
\safemath{\bmis}{\bisd}
\safemath{\bmit}{\bitd}
\safemath{\bmiu}{\biud}
\safemath{\bmiv}{\bivd}
\safemath{\bmiw}{\biwd}
\safemath{\bmix}{\bixd}
\safemath{\bmiy}{\biyd}
\safemath{\bmiz}{\bizd}

\safemath{\bmxi}{\bixid}
\safemath{\bmlambda}{\bilambdad}
\safemath{\bmmu}{\bimud}
\safemath{\bmtheta}{\bithetad}
\safemath{\bmphi}{\biphid}
\safemath{\bmdelta}{\bideltad}

\safemath{\bA}{\mathbf{A}}
\safemath{\bB}{\mathbf{B}}
\safemath{\bC}{\mathbf{C}}
\safemath{\bD}{\mathbf{D}}
\safemath{\bE}{\mathbf{E}}
\safemath{\bF}{\mathbf{F}}
\safemath{\bG}{\mathbf{G}}
\safemath{\bH}{\mathbf{H}}
\safemath{\bI}{\mathbf{I}}
\safemath{\bJ}{\mathbf{J}}
\safemath{\bK}{\mathbf{K}}
\safemath{\bL}{\mathbf{L}}
\safemath{\bM}{\mathbf{M}}
\safemath{\bN}{\mathbf{N}}
\safemath{\bO}{\mathbf{O}}
\safemath{\bP}{\mathbf{P}}
\safemath{\bQ}{\mathbf{Q}}
\safemath{\bR}{\mathbf{R}}
\safemath{\bS}{\mathbf{S}}
\safemath{\bT}{\mathbf{T}}
\safemath{\bU}{\mathbf{U}}
\safemath{\bV}{\mathbf{V}}
\safemath{\bW}{\mathbf{W}}
\safemath{\bX}{\mathbf{X}}
\safemath{\bY}{\mathbf{Y}}
\safemath{\bZ}{\mathbf{Z}}

\safemath{\bZero}{\mathbf{0}}
\safemath{\bOne}{\mathbf{1}}
\safemath{\bDelta}{\mathbf{\Delta}}
\safemath{\bLambda}{\mathbf{\UpLambda}}
\safemath{\bPhi}{\mathbf{\Upphi}}
\safemath{\bSigma}{\mathbf{\Upsigma}}
\safemath{\bOmega}{\mathbf{\Upomega}}
\safemath{\bTheta}{\mathbf{\Uptheta}}

\bmdefine{\biAd}{A}
\bmdefine{\biBd}{B}
\bmdefine{\biCd}{C}
\bmdefine{\biDd}{D}
\bmdefine{\biEd}{E}
\bmdefine{\biFd}{F}
\bmdefine{\biGd}{G}
\bmdefine{\biHd}{H}
\bmdefine{\biId}{I}
\bmdefine{\biJd}{J}
\bmdefine{\biKd}{K}
\bmdefine{\biLd}{L}
\bmdefine{\biMd}{M}
\bmdefine{\biOd}{N}
\bmdefine{\biPd}{O}
\bmdefine{\biQd}{P}
\bmdefine{\biRd}{R}
\bmdefine{\biSd}{S}
\bmdefine{\biTd}{T}
\bmdefine{\biUd}{U}
\bmdefine{\biVd}{V}
\bmdefine{\biWd}{W}
\bmdefine{\biXd}{X}
\bmdefine{\biYd}{Y}
\bmdefine{\biZd}{Z}

\bmdefine{\biDelta}{\Delta}
\bmdefine{\biLambda}{\Lambda}
\bmdefine{\biPhi}{\Phi}
\bmdefine{\biSigma}{\Sigma}
\bmdefine{\biOmega}{\Omega}
\bmdefine{\biTheta}{\Theta}

\safemath{\bimA}{\biAd}
\safemath{\bimB}{\biBd}
\safemath{\bimC}{\biCd}
\safemath{\bimD}{\biDd}
\safemath{\bimE}{\biEd}
\safemath{\bimF}{\biFd}
\safemath{\bimG}{\biGd}
\safemath{\bimH}{\biHd}
\safemath{\bimI}{\biId}
\safemath{\bimJ}{\biJd}
\safemath{\bimK}{\biKd}
\safemath{\bimL}{\biLd}
\safemath{\bimM}{\biMd}
\safemath{\bimN}{\biNd}
\safemath{\bimO}{\biOd}
\safemath{\bimP}{\biPd}
\safemath{\bimQ}{\biQd}
\safemath{\bimR}{\biRd}
\safemath{\bimS}{\biSd}
\safemath{\bimT}{\biTd}
\safemath{\bimU}{\biUd}
\safemath{\bimV}{\biVd}
\safemath{\bimW}{\biWd}
\safemath{\bimX}{\biXd}
\safemath{\bimY}{\biYd}
\safemath{\bimZ}{\biZd}

\safemath{\bimDelta}{\biDelta}
\safemath{\bimLambda}{\biLambda}
\safemath{\bimPhi}{\biPhi}
\safemath{\bimSigma}{\biSigma}
\safemath{\bimOmega}{\biOmega}
\safemath{\bimTheta}{\biTheta}

\safemath{\setA}{\mathcal{A}}
\safemath{\setB}{\mathcal{B}}
\safemath{\setC}{\mathcal{C}}
\safemath{\setD}{\mathcal{D}}
\safemath{\setE}{\mathcal{E}}
\safemath{\setF}{\mathcal{F}}
\safemath{\setG}{\mathcal{G}}
\safemath{\setH}{\mathcal{H}}
\safemath{\setI}{\mathcal{I}}
\safemath{\setJ}{\mathcal{J}}
\safemath{\setK}{\mathcal{K}}
\safemath{\setL}{\mathcal{L}}
\safemath{\setM}{\mathcal{M}}
\safemath{\setN}{\mathcal{N}}
\safemath{\setO}{\mathcal{O}}
\safemath{\setP}{\mathcal{P}}
\safemath{\setQ}{\mathcal{Q}}
\safemath{\setR}{\mathcal{R}}
\safemath{\setS}{\mathcal{S}}
\safemath{\setT}{\mathcal{T}}
\safemath{\setU}{\mathcal{U}}
\safemath{\setV}{\mathcal{V}}
\safemath{\setW}{\mathcal{W}}
\safemath{\setX}{\mathcal{X}}
\safemath{\setY}{\mathcal{Y}}
\safemath{\setZ}{\mathcal{Z}}
\safemath{\emptySet}{\varnothing}

\safemath{\colA}{\mathscr{A}}
\safemath{\colB}{\mathscr{B}}
\safemath{\colC}{\mathscr{C}}
\safemath{\colD}{\mathscr{D}}
\safemath{\colE}{\mathscr{E}}
\safemath{\colF}{\mathscr{F}}
\safemath{\colG}{\mathscr{G}}
\safemath{\colH}{\mathscr{H}}
\safemath{\colI}{\mathscr{I}}
\safemath{\colJ}{\mathscr{J}}
\safemath{\colK}{\mathscr{K}}
\safemath{\colL}{\mathscr{L}}
\safemath{\colM}{\mathscr{M}}
\safemath{\colN}{\mathscr{N}}
\safemath{\colO}{\mathscr{O}}
\safemath{\colP}{\mathscr{P}}
\safemath{\colQ}{\mathscr{Q}}
\safemath{\colR}{\mathscr{R}}
\safemath{\colS}{\mathscr{S}}
\safemath{\colT}{\mathscr{T}}
\safemath{\colU}{\mathscr{U}}
\safemath{\colV}{\mathscr{V}}
\safemath{\colW}{\mathscr{W}}
\safemath{\colX}{\mathscr{X}}
\safemath{\colY}{\mathscr{Y}}
\safemath{\colZ}{\mathscr{Z}}

\safemath{\opA}{\mathbb{A}}
\safemath{\opB}{\mathbb{B}}
\safemath{\opC}{\mathbb{C}}
\safemath{\opD}{\mathbb{D}}
\safemath{\opE}{\mathbb{E}}
\safemath{\opF}{\mathbb{F}}
\safemath{\opG}{\mathbb{G}}
\safemath{\opH}{\mathbb{H}}
\safemath{\opI}{\mathbb{I}}
\safemath{\opJ}{\mathbb{J}}
\safemath{\opK}{\mathbb{K}}
\safemath{\opL}{\mathbb{L}}
\safemath{\opM}{\mathbb{M}}
\safemath{\opN}{\mathbb{N}}
\safemath{\opO}{\mathbb{O}}
\safemath{\opP}{\mathbb{P}}
\safemath{\opQ}{\mathbb{Q}}
\safemath{\opR}{\mathbb{R}}
\safemath{\opS}{\mathbb{S}}
\safemath{\opT}{\mathbb{T}}
\safemath{\opU}{\mathbb{U}}
\safemath{\opV}{\mathbb{V}}
\safemath{\opW}{\mathbb{W}}
\safemath{\opX}{\mathbb{X}}
\safemath{\opY}{\mathbb{Y}}
\safemath{\opZ}{\mathbb{Z}}
\safemath{\opZero}{\mathbb{O}}
\safemath{\identityop}{\opI}

\safemath{\veca}{\bma}
\safemath{\vecb}{\bmb}
\safemath{\vecc}{\bmc}
\safemath{\vecd}{\bmd}
\safemath{\vece}{\bme}
\safemath{\vecf}{\bmf}
\safemath{\vecg}{\bmg}
\safemath{\vech}{\bmh}
\safemath{\veci}{\bmi}
\safemath{\vecj}{\bmj}
\safemath{\veck}{\bmk}
\safemath{\vecl}{\bml}
\safemath{\vecm}{\bmm}
\safemath{\vecn}{\bmn}
\safemath{\veco}{\bmo}
\safemath{\vecp}{\bmp}
\safemath{\vecq}{\bmq}
\safemath{\vecr}{\bmr}
\safemath{\vecs}{\bms}
\safemath{\vect}{\bmt}
\safemath{\vecu}{\bmu}
\safemath{\vecv}{\bmv}
\safemath{\vecw}{\bmw}
\safemath{\vecx}{\bmx}
\safemath{\vecy}{\bmy}
\safemath{\vecz}{\bmz}

\safemath{\veczero}{\bmzero}
\safemath{\vecone}{\bmone}
\safemath{\vecxi}{\bmxi}
\safemath{\veclambda}{\bmlambda}
\safemath{\vecmu}{\bmmu}
\safemath{\vectheta}{\bmtheta}
\safemath{\vecphi}{\bmphi}
\safemath{\vecdelta}{\bmdelta}

\safemath{\matA}{\bA}
\safemath{\matB}{\bB}
\safemath{\matC}{\bC}
\safemath{\matD}{\bD}
\safemath{\matE}{\bE}
\safemath{\matF}{\bF}
\safemath{\matG}{\bG}
\safemath{\matH}{\bH}
\safemath{\matI}{\bI}
\safemath{\matJ}{\bJ}
\safemath{\matK}{\bK}
\safemath{\matL}{\bL}
\safemath{\matM}{\bM}
\safemath{\matN}{\bN}
\safemath{\matO}{\bO}
\safemath{\matP}{\bP}
\safemath{\matQ}{\bQ}
\safemath{\matR}{\bR}
\safemath{\matS}{\bS}
\safemath{\matT}{\bT}
\safemath{\matU}{\bU}
\safemath{\matV}{\bV}
\safemath{\matW}{\bW}
\safemath{\matX}{\bX}
\safemath{\matY}{\bY}
\safemath{\matZ}{\bZ}
\safemath{\matzero}{\bmzero}

\safemath{\matDelta}{\bDelta}
\safemath{\matLambda}{\bLambda}
\safemath{\matPhi}{\bPhi}
\safemath{\matSigma}{\bSigma}
\safemath{\matOmega}{\bOmega}
\safemath{\matTheta}{\bTheta}

\safemath{\matidentity}{\matI}
\safemath{\matone}{\matO}

\safemath{\rnda}{A}
\safemath{\rndb}{B}
\safemath{\rndc}{C}
\safemath{\rndd}{D}
\safemath{\rnde}{E}
\safemath{\rndf}{F}
\safemath{\rndg}{G}
\safemath{\rndh}{H}
\safemath{\rndi}{I}
\safemath{\rndj}{J}
\safemath{\rndk}{K}
\safemath{\rndl}{L}
\safemath{\rndm}{M}
\safemath{\rndn}{N}
\safemath{\rndo}{O}
\safemath{\rndp}{P}
\safemath{\rndq}{Q}
\safemath{\rndr}{R}
\safemath{\rnds}{S}
\safemath{\rndt}{T}
\safemath{\rndu}{U}
\safemath{\rndv}{V}
\safemath{\rndw}{W}
\safemath{\rndx}{X}
\safemath{\rndy}{Y}
\safemath{\rndz}{Z}

\safemath{\rveca}{\bimA}
\safemath{\rvecb}{\bimB}
\safemath{\rvecc}{\bimC}
\safemath{\rvecd}{\bimD}
\safemath{\rvece}{\bimE}
\safemath{\rvecf}{\bimF}
\safemath{\rvecg}{\bimG}
\safemath{\rvech}{\bimH}
\safemath{\rveci}{\bimI}
\safemath{\rvecj}{\bimJ}
\safemath{\rveck}{\bimK}
\safemath{\rvecl}{\bimL}
\safemath{\rvecm}{\bimM}
\safemath{\rvecn}{\bimN}
\safemath{\rveco}{\bomO}
\safemath{\rvecp}{\bimP}
\safemath{\rvecq}{\bimQ}
\safemath{\rvecr}{\bimR}
\safemath{\rvecs}{\bimS}
\safemath{\rvect}{\bimT}
\safemath{\rvecu}{\bimU}
\safemath{\rvecv}{\bimV}
\safemath{\rvecw}{\bimW}
\safemath{\rvecx}{\bimX}
\safemath{\rvecy}{\bimY}
\safemath{\rvecz}{\bimZ}

\safemath{\rvecxi}{\bmxi}
\safemath{\rveclambda}{\bmlambda}
\safemath{\rvecmu}{\bmmu}
\safemath{\rvectheta}{\bmtheta}
\safemath{\rvecphi}{\bmphi}

\safemath{\rmatA}{\bimA}
\safemath{\rmatB}{\bimB}
\safemath{\rmatC}{\bimC}
\safemath{\rmatD}{\bimD}
\safemath{\rmatE}{\bimE}
\safemath{\rmatF}{\bimF}
\safemath{\rmatG}{\bimG}
\safemath{\rmatH}{\bimH}
\safemath{\rmatI}{\bimI}
\safemath{\rmatJ}{\bimJ}
\safemath{\rmatK}{\bimK}
\safemath{\rmatL}{\bimL}
\safemath{\rmatM}{\bimM}
\safemath{\rmatN}{\bimN}
\safemath{\rmatO}{\bimO}
\safemath{\rmatP}{\bimP}
\safemath{\rmatQ}{\bimQ}
\safemath{\rmatR}{\bimR}
\safemath{\rmatS}{\bimS}
\safemath{\rmatT}{\bimT}
\safemath{\rmatU}{\bimU}
\safemath{\rmatV}{\bimV}
\safemath{\rmatW}{\bimW}
\safemath{\rmatX}{\bimX}
\safemath{\rmatY}{\bimY}
\safemath{\rmatZ}{\bimZ}

\safemath{\rmatDelta}{\bimDelta}
\safemath{\rmatLambda}{\bimLambda}
\safemath{\rmatPhi}{\bimPhi}
\safemath{\rmatSigma}{\bimSigma}
\safemath{\rmatOmega}{\bimOmega}
\safemath{\rmatTheta}{\bimTheta}

\usepackage{amssymb}
\usepackage{amsfonts}
\usepackage{mathrsfs}
\usepackage{xspace}
\usepackage{bm}
\usepackage{fancyref}
\usepackage{textcomp}

\usepackage{multirow}
\usepackage{stmaryrd}

\newenvironment{textbmatrix}{	\setlength{\arraycolsep}{2.5pt}%
								\big[\begin{matrix}}{\end{matrix}\big]%
								\raisebox{0.08ex}{\vphantom{M}}}

\def\be{\begin{equation}}
\def\ee{\end{equation}}
\def\een{\nonumber \end{equation}}
\def\mat{\begin{bmatrix}}
\def\emat{\end{bmatrix}}
\def\btm{\begin{textbmatrix}}
\def\etm{\end{textbmatrix}}

\def\ba#1\ea{\begin{align}#1\end{align}}
\def\bas#1\eas{\begin{align*}#1\end{align*}}
\def\bs#1\es{\begin{split}#1\end{split}} 
\def\bg#1\eg{\begin{gather}#1\end{gather}}
\def\bml#1\eml{\begin{multline}#1\end{multline}}
\def\bi#1\ei{\begin{itemize}#1\end{itemize}}

\newcommand{\subtext}[1]{\text{\fontfamily{cmr}\fontshape{n}\fontseries{m}\selectfont{}#1}}
\newcommand{\lefto}{\mathopen{}\left}

\DeclareMathOperator{\sign}{sgn}			
\DeclareMathOperator*{\argmin}{arg\;min}		
\DeclareMathOperator{\Exop}{\opE}			
\DeclareMathOperator{\Varop}{\mathrm{Var}}
\DeclareMathOperator{\Covop}{\mathrm{Cov}}
\DeclareMathOperator{\rot}{\nabla\times}		

\newcommand{\Ex}[2]{\ensuremath{\Exop_{#1}\lefto[#2\right]}} 	
\newcommand{\abs}[1]{\lefto\lvert#1\right\rvert}		

\newcommand{\vecnorm}[1]{\lefto\lVert#1\right\rVert}		
\newcommand{\opnorm}[1]{\lVert#1\rVert}		

\safemath{\dirac}{\delta}					
\safemath{\krond}{\dirac}					

\safemath{\upto}{\uparrow}
\safemath{\downto}{\downarrow}
\safemath{\iu}{j}							
\safemath{\ev}{\lambda}						
\safemath{\hilseqspace}{l^{2}}				
\newcommand{\banachfunspace}[1]{\setL^{#1}}	
\safemath{\hilfunspace}{\banachfunspace{2}}	
\newcommand{\floor}[1]{\lfloor #1 \rfloor}

\safemath{\SNR}{\textsf{SNR}} 				
\safemath{\PAR}{\textsf{PAR}} 				
\safemath{\No}{N_0}							
\safemath{\Es}{E_s}							
\safemath{\Eb}{E_b}							
\safemath{\EbNo}{\frac{\Eb}{\No}}
\safemath{\EsNo}{\frac{\Es}{\No}}

\DeclareMathOperator{\CHop}{\ensuremath{\opH}} 
\safemath{\tvir}{\rndh_{\CHop}}				
\safemath{\tvtf}{\rndl_{\CHop}}				
\safemath{\spf}{\rnds_{\CHop}}				
\safemath{\bff}{H_{\CHop}}					

\safemath{\ircf}{r_{h}}						
\safemath{\tftvcf}{r_{s}}					
\safemath{\tfcf}{r_{l}}						
\safemath{\bfcf}{r_{H}}						

\safemath{\tcorr}{c_h}						
\safemath{\scf}{c_{s}}						
\safemath{\tfcorr}{c_{l}}					
\safemath{\fcorr}{c_{H}}						

\safemath{\mi}{I}							
\safemath{\capacity}{C}						

\safemath{\normal}{\mathcal{N}}			
\safemath{\jpg}{\mathcal{CN}}			
\safemath{\mchain}{\leftrightarrow}		
\newcommand{\AIC}{\ensuremath{\text{AIC}}} 

\safemath{\dB}{\,\mathrm{dB}}
\safemath{\dBm}{\,\mathrm{dBm}}
\safemath{\Hz}{\,\mathrm{Hz}}
\safemath{\kHz}{\,\mathrm{kHz}}
\safemath{\MHz}{\,\mathrm{MHz}}
\safemath{\GHz}{\,\mathrm{GHz}}
\safemath{\s}{\,\mathrm{s}}
\safemath{\ms}{\,\mathrm{ms}}
\safemath{\mus}{\,\mathrm{\text{\textmu}s}}
\safemath{\ns}{\,\mathrm{ns}}
\safemath{\ps}{\,\mathrm{ps}}
\safemath{\meter}{\,\mathrm{m}}
\safemath{\mm}{\,\mathrm{mm}}
\safemath{\cm}{\,\mathrm{cm}}
\safemath{\m}{\,\mathrm{m}}
\safemath{\W}{\,\mathrm{W}}
\safemath{\mW}{\, \mathrm{mW}}
\safemath{\J}{\,\mathrm{J}}
\safemath{\K}{\,\mathrm{K}}
\safemath{\bit}{\,\mathrm{bit}}
\safemath{\nat}{\,\mathrm{nat}}

\safemath{\define}{\triangleq}			

\safemath{\equivalent}{\sim}
\safemath{\distas}{\sim}					
\safemath{\sdiff}{\Delta}				

\safemath{\reals}{\mathbb{R}}
\safemath{\positivereals}{\reals_{+}}
\safemath{\integers}{\mathbb{Z}}
\safemath{\posint}{\integers_{+}}
\safemath{\naturals}{\mathbb{N}}
\safemath{\posnaturals}{\naturals_{+}}
\safemath{\complexset}{\mathbb{C}}
\safemath{\rationals}{\mathbb{Q}}

\newcommand*{\fancyrefapplabelprefix}{app}		
\newcommand*{\fancyrefthmlabelprefix}{thm}		
\newcommand*{\fancyreflemlabelprefix}{lem}		
\newcommand*{\fancyrefcorlabelprefix}{cor}		
\newcommand*{\fancyrefdeflabelprefix}{def}		
\newcommand*{\fancyrefproplabelprefix}{prop}	
\newcommand*{\fancyrefobslabelprefix}{obs}		
\newcommand*{\fancyrefalglabelprefix}{alg}		
\newcommand*{\fancyrefasmlabelprefix}{asm}	    
\newcommand*{\fancyreftbllabelprefix}{tab}	 

\frefformat{vario}{\fancyrefseclabelprefix}{Section~#1}
\frefformat{vario}{\fancyrefthmlabelprefix}{Theorem~#1}
\frefformat{vario}{\fancyreflemlabelprefix}{Lemma~#1}
\frefformat{vario}{\fancyrefcorlabelprefix}{Corollary~#1}
\frefformat{vario}{\fancyrefdeflabelprefix}{Definition~#1}
\frefformat{vario}{\fancyrefobslabelprefix}{Observation~#1}
\frefformat{vario}{\fancyrefasmlabelprefix}{Assumption~#1}
\frefformat{vario}{\fancyreffiglabelprefix}{Fig.~#1}
\frefformat{vario}{\fancyrefapplabelprefix}{Appendix~#1} 
\frefformat{vario}{\fancyrefproplabelprefix}{Proposition~#1}
\frefformat{vario}{\fancyrefalglabelprefix}{Algorithm~#1}
\frefformat{vario}{\fancyreftbllabelprefix}{Table~#1}
\frefformat{vario}{\fancyrefeqlabelprefix}{(#1)}

\newcommand{\marginnote}[1]{\marginpar{\framebox{\framebox{#1}}}}
\safemath{\dictab}{[\,\dicta\,\,\dictb\,]}

\safemath{\ysig}{\bmy}
\safemath{\ysighat}{\hat{\ysig}}
\safemath{\ysigdim}{M}
\safemath{\xsig}{\bmx}
\safemath{\xsigdim}{N}
\safemath{\nx}{n_x}
\safemath{\zsig}{\bmz}
\safemath{\zsigdim}{\ysigdim}
\safemath{\rsig}{\bmr}
\safemath{\Adict}{\bA}
\safemath{\Adicttilde}{\widetilde{\Adict}}
\safemath{\Adictdim}{\outputdim\times\xsigdim}
\safemath{\avec}{\bma}
\safemath{\avectilde}{\tilde{\avec}}
\safemath{\Bdict}{\bB}
\safemath{\Bdicttilde}{\widetilde{\Bdict}}
\safemath{\Cdict}{\bC}
\safemath{\cvec}{\bmc}
\safemath{\Ddict}{\bD}
\safemath{\Ddictdim}{\ysigdim\times\xsigdim}
\safemath{\dvec}{\bmd}
\safemath{\Ddicttilde}{\widetilde{\bD}}
\safemath{\Bonb}{\bB}
\safemath{\bvec}{\bmb}
\safemath{\Bonbdim}{\ysigdim\times\ysigdim}
\safemath{\noise}{\bmn}
\safemath{\noisedim}{\ysigim}
\safemath{\err}{\bme}
\safemath{\errdim}{\ysigdim}
\safemath{\errset}{\setE}
\safemath{\nerr}{n_e}
\safemath{\delop}{\bP_\errset}
\safemath{\delopc}{\bP_{{\errset}^c}}

\safemath{\cplxi}{\imath}
\safemath{\cplxj}{\jmath}

\safemath{\dict}{\matD}
\safemath{\inputdim}{N}		
\safemath{\outputdim}{M}		
\safemath{\sparsity}{S}	
\safemath{\inputdimA}{{N_a}}	
\safemath{\inputdimB}{{N_b}}	
\safemath{\elemA}{{n_a}}	
\safemath{\elemB}{{n_b}}	
\safemath{\resA}{\matR_a}	
\safemath{\resB}{\matR_b}	
\safemath{\subD}{\matS} 
\safemath{\subA}{\matS_a} 
\safemath{\subB}{\matS_b} 
\safemath{\dicta}{\matA} 	
\safemath{\dictb}{\matB} 	
\safemath{\hollowS}{H}
\safemath{\hollowA}{H_a}
\safemath{\hollowB}{H_b}
\safemath{\cross}{Z}
\safemath{\coh}{\mu_d}			
\safemath{\coha}{\mu_a}			
\safemath{\cohb}{\mu_b}			
\safemath{\mubs}{\nu}	
\safemath{\cohm}{\mu_m} 
\safemath{\dictset}{\setD}	
\safemath{\dictsetp}{\dictset(\coh,\coha,\cohb)}	
\safemath{\dictsetgen}{\dictset_\text{gen}}
\safemath{\dictsetgenp}{\dictsetgen(\coh)}
\safemath{\dictsetonb}{\dictset_\text{onb}}
\safemath{\dictsetonbp}{\dictsetonb(\coh)}

\safemath{\leftside}{U}
\safemath{\rightsideA}{R_a}
\safemath{\rightsideB}{R_b}

\safemath{\indexS}{\setI_S} 

\safemath{\na}{n_a}			
\safemath{\nb}{n_b}			
\safemath{\coeffa}{p_i}	
\safemath{\coeffb}{q_j}	
\safemath{\seta}{\setP}		
\safemath{\setb}{\setQ}     
\safemath{\setw}{\setW}	
\safemath{\setz}{\setZ}	
\safemath{\cola}{\veca}		
\safemath{\colb}{\vecb}		
\safemath{\cold}{\vecd}		
\safemath{\inputvec}{\vecx} 	
\safemath{\error}{\vece}	
\safemath{\noiseout}{\vecz} 	
\safemath{\inputvecel}{x}
\safemath{\inputveca}{\vecx_a}
\safemath{\inputvecb}{\vecx_b}
\safemath{\outputvec}{\vecy}	
\safemath{\lambdamin}{\lambda_{\mathrm{min}}}

\safemath{\elltwo}{\ell_2}
\safemath{\ellone}{\ell_1}
\safemath{\ellzero}{\ell_0}
\safemath{\ellinf}{\ell_\infty}
\safemath{\ellinftilde}{\ell_{\widetilde\infty}}
\safemath{\licard}{Z(\coh,\coha,\cohb)}
\safemath{\xsol}{\hat{x}}
\safemath{\xbord}{x_b}		
\safemath{\xstat}{x_s}		
\safemath{\xstatLone}{\tilde{x}_s}
\safemath{\order}{\mathcal{O}} 
\safemath{\scales}{\Theta} 
\safemath{\ones}{\mathbf{1}} 
\safemath{\zeroes}{\mathbf{0}} 
\safemath{\thlone}{\kappa(\coh,\cohb)} 
\safemath{\constoneA}{\delta} 
\safemath{\constoneB}{\epsilon} 
\safemath{\nlarge}{L}				   
\safemath{\sumlarge}{S_\nlarge}
\safemath{\maxlarger}{P_\nlarge}	   
\safemath{\Pzero}{\textrm{P0}}	
\safemath{\Pone}{\textrm{P1}}
\safemath{\vecfir}{\vecw}			 
\safemath{\vecsec}{\vecz}
\safemath{\elvecfir}{w}              
\safemath{\elvecsec}{z}				 
\safemath{\nlargefir}{n}
\safemath{\normout}{\gamma}
\safemath{\auxfun}{h}
\safemath{\supp}{\textrm{supp}}

\safemath{\indexa}{\ell}
\safemath{\indexb}{r}
\safemath{\indexc}{i}
\safemath{\indexd}{j}

\safemath{\project}{P}

\IEEEoverridecommandlockouts
\allowdisplaybreaks

\begin{document}

%
\title{Nonlinear Precoding for Phase-Quantized Constant-Envelope Massive MU-MIMO-OFDM}
\author{\IEEEauthorblockN{Sven Jacobsson$^\text{1,2}$, Oscar Casta\~{n}eda$^\text{3}$, Charles Jeon$^\text{3}$, Giuseppe Durisi$^\text{1}$, and Christoph Studer$^\text{3}$}
\thanks{The work of SJ and GD was supported in part by the Swedish Foundation for Strategic Research under grant ID14-0022, and by the Swedish Governmental Agency for Innovation Systems (VINNOVA) within the competence center ChaseOn. SJ's research visit at Cornell was sponsored in part by Cornell's College of Engineering. The work of OC, CJ, and CS was supported by Xilinx, Inc.~and by the US National Science Foundation~(NSF) under grants ECCS-1408006, CCF-1535897, CAREER CCF-1652065, and CNS-1717559.}
\IEEEauthorblockA{
$^\text{1}$Chalmers University of Technology, Gothenburg, Sweden\\
$^\text{2}$Ericsson Research, Gothenburg, Sweden\\
$^\text{3}$Cornell University, Ithaca, NY, USA
}}
\maketitle

\begin{abstract}

We propose a nonlinear phase-quantized constant-envelope precoding algorithm for the massive multi-user (MU) multiple-input multiple-output (MIMO) downlink. Specifically, we adapt the squared-infinity norm Douglas-Rachford splitting (SQUID) precoder to systems that use oversampling digital-to-analog converters~(DACs) at the base station~(BS) and orthogonal frequency-division multiplexing (OFDM) to communicate over frequency-selective channels.
We demonstrate that the proposed SQUID-OFDM precoder is able to generate transmit signals that are constrained to constant envelope, which enables the use of power-efficient analog radio-frequency circuitry at the BS. By quantizing the phase of the resulting constant-envelope signal, we obtain a finite-cardinality transmit signal that can be synthesized by low-resolution (e.g., 1-bit) DACs. 
We use error-rate simulations to demonstrate the superiority of SQUID-OFDM over linear-quantized precoders for massive MU-MIMO-OFDM~systems. 

\end{abstract}

\section{Introduction}

Massive \ac{MU} \ac{MIMO} equips the \ac{BS} with a large number of antenna elements and serves tens of \acp{UE} simultaneously and in the same frequency band~\cite{boccardi14a, larsson14a}.
While massive  \ac{MU}-\ac{MIMO} is expected to be a key technology component of fifth-generation (5G) wireless networks, scaling traditional \ac{RF} front-end architectures to \acp{BS} with hundreds of antenna elements leads to a prohibitive growth in circuit power consumption, system costs, and hardware complexity. 
Hence, a successful deployment of massive MU-MIMO requires inexpensive, power-efficient, and low-complexity hardware components, which, in turn, will limit the capacity of the system due to signal-quality degradation.

\subsection{Constant-Envelope and Phase-Quantized Precoding}

In the massive \ac{MU}-\ac{MIMO} downlink (the \ac{BS} transmits data to the \acp{UE}), precoding must be used to reduce \ac{MU} interference. Unfortunately, precoding typically generates time-domain signals with high \ac{PAR}~\cite{mollen16e}; this fact is further aggravated in systems that use \ac{OFDM} to facilitate communication over wideband frequency-selective channels~\cite{han05a}. 
For such high-\ac{PAR} waveforms, one has to operate the \acp{PA} in the linear regime to prevent significant signal-quality degradation. This results in high \ac{PA} power consumption~\cite{han05a}.
 
\setlength{\textfloatsep}{10pt}
\begin{figure*}
\centering
 \includegraphics[width=0.75\textwidth]{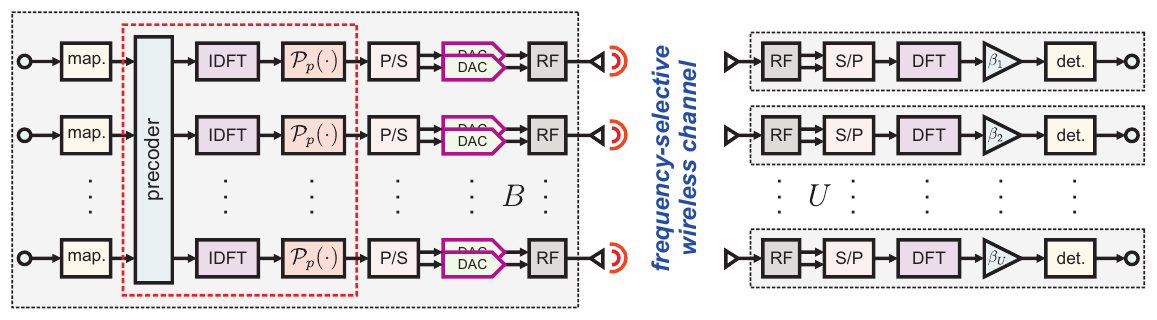}
 \vspace{-0.1cm}
 \caption{Overview of the considered massive MU-MIMO-OFDM downlink system. Left: BS with $B$ antennas performs precoding in the frequency domain, transforms the precoded vector into time domain, and maps its entries to the set of outcomes supported by the   transcoder in the DACs. The dashed red box indicates the operations carried out by the nonlinear phase-quantized constant-envelope precoder. Right: $U$ single-antenna~UEs. } \label{fig:overview}
\vspace{-0.1cm}
\end{figure*}

%
To mitigate the high-\ac{PAR} issue, a constant-envelope precoder for the massive \ac{MU}-\ac{MIMO}-\ac{OFDM} case was proposed in~\cite{mohammed13b}; its design ensures that the precoded signal has equal amplitude on all antennas (and, hence, zero PAR). This precoder enables \acp{PA} to operate in the nonlinear regime, allowing for energy-efficient analog circuitry.
Recently, the authors of \cite{noll17a} designed a precoder  for the frequency-flat case that outputs a constant-envelope signal constrained to only eight phases.
This precoder requires the \acp{DAC} at the \ac{BS} to generate only eight phase outputs, which enables the use of power-efficient converter architectures, and reduces the interconnect data rates between the baseband-processing unit and the radio~unit at the BS.

\subsection{1-Bit Precoding}

Motivated by potential power savings and reduced interconnect data rates, the use of 1-bit \acp{DAC} in the massive MU-MIMO downlink has recently attracted significant attention.
Specifically, so-called \emph{linear-quantized} precoders (i.e., linear precoding followed by quantization) have been recently proposed  for precoding in massive MU-MIMO-OFDM systems that use oversampling DACs~\cite{jacobsson17e,jacobsson17c}. These precoders achieve low \acp{BER} and high sum-rate throughputs over frequency-selective channels with \ac{OFDM}, despite the adverse impact of the 1-bit~\acp{DAC}.
\emph{Nonlinear} precoders, where the precoder depends on the instantaneous realizations of the information symbols, are known to significantly outperform linear-quantized precoders~(see, e.g.,~\cite{jedda16a, swindlehurst17a, jacobsson17d, jacobsson16d, castaneda17a, li17c, landau17a}), but have, until recently, been analyzed exclusively for frequency-flat channels and single-carrier transmission.\footnote{See, however,~\cite{nedelcu17a} for a recent result on the frequency-selective case with OFDM transmission.}

\subsection{Contributions}

We propose a nonlinear phase-quantized constant-envelope precoder for the massive \ac{MU}-\ac{MIMO}-\ac{OFDM} downlink operating over frequency-selective channels. Our precoder builds upon the \ac{SQUID} algorithm put forward in~\cite[Sec.~IV-B]{jacobsson17d}. 
In contrast to previous works~\cite{jacobsson17d, jedda16a, swindlehurst17a, jacobsson16d, castaneda17a, li17c, landau17a}, which focus on the case of Nyquist-rate sampling 1-bit \acp{DAC} and on the frequency-flat case, the proposed nonlinear precoder, which we shall refer to as {SQUID-OFDM}, is capable of supporting oversampling \acp{DAC} and~\ac{OFDM}. 
We characterize the computational complexity of SQUID-OFDM and demonstrate its efficacy via numerical~simulations.

\subsection{Notation}
Lowercase and uppercase boldface letters denote vectors and matrices, respectively. 
%
%
The $M \times N$ all-zeros matrix and the $M \times M$ identity matrix are denoted by $\veczero_{M \times N}$ and $\matI_M$, respectively.
The unitary $N \times N$ \ac{DFT} matrix is denoted by~$\matF_N$. 
The $\ell_\infty$-norm of~$\veca = [a_1, \dots, a_M]^T$ is $\opnorm{\veca}_\infty = \max_{\ell=1,\ldots,M}  \abs{a_\ell}$; the $\ell_{\widetilde\infty}$-norm  is $\opnorm{\veca}_{\widetilde\infty} = \max\big\{ \|\Re\{\bma\}\|_\infty, \|\Im\{\bma\}\|_\infty \big\} $.
We use $\opnorm{\veca}_2$ and $\opnorm{\matA}_F$ to denote the $\ell_2$-norm of  vector $\veca$ and the Frobenius norm of matrix~$\matA$, respectively. If $\matA$ is an $M \times N$ matrix, then $\text{vec}(\matA)$ is an $MN$-dimensional vector obtained by column-wise vectorization of $\matA$. 
The phase of $a \in \opC$ is denoted by $\arg(a)$; the sign of~$r \in \opR$ is denoted by $\sign(r) \in \{ -1, +1\}$.
The floor function $\floor{r}$ produces the largest integer less than or equal to $r$.
%
The complex-valued circularly symmetric Gaussian distribution with covariance matrix $\matK \in \opC^{M \times M}$ is denoted by~$\jpg(\matzero_{M \times 1}, \matK)$.
The expected value of $\matA$ is~$\Ex{}{\matA}$.

\section{System Model} \label{sec:system}

We consider a single-cell massive MU-MIMO-OFDM downlink system as illustrated in Fig.~\ref{fig:overview}. The system operates over a wideband channel where \ac{OFDM} is used to deal with the selectiveness in frequency of the channel.
Let $B$ denote the number of \ac{BS} antennas and  $U$  the number of single-antenna \acp{UE}.
At the BS, the frequency-domain information symbols are mapped to the antenna array by a precoder. At each BS antenna, the precoded signal is mapped to time domain through an \ac{IDFT} before being passed to a pair of finite-resolution \acp{DAC}, which  generate the in-phase and quadrature components of the transmitted time-domain signal.
For simplicity, we ignore other \ac{RF} impairments and assume perfect synchronization between the \ac{BS} and the \acp{UE}.

\subsection{Channel Input-Output Relation}
Under the above assumptions, the received signal $\vecy_n \in \opC^U$ at the $U$ \acp{UE} can be written as
\begin{IEEEeqnarray}{rCl} \label{eq:received_freq}
	\vecy_n &=& \sum_{\ell = 0}^{L-1} \matH_{\ell} \vecx_{n-\ell} + \vecw_n
\end{IEEEeqnarray}
at discrete time instants $n=0,\ldots,N-1$. Here, $\vecx_n$ is the $B$-dimensional transmit signal at discrete time~$n$ and $N$ is the number of samples per OFDM symbol (the size of the \ac{IDFT}). 
The vector $\vecw_n \sim \jpg(\veczero_{U \times 1}, N_0 \matI_U)$ denotes the i.i.d.\ \ac{AWGN} at the \acp{UE} at discrete time $n$. Here,~$N_0$ is the noise power and $\textit{SNR} = 1/N_0$ defines the \ac{SNR}. The matrix $\matH_\ell \in \opC^{U \times B}$ is the $\ell$th tap of the frequency-selective channel ($\ell = 0, \dots, L-1$). We assume that the realizations of~$\{ \matH_\ell \}_{\ell = 0}^{L-1}$ remain constant for the duration of each \ac{OFDM} symbol and that they are perfectly known to the~\ac{BS}.
Let $\matX = [\vecx_0,\dots,\vecx_{N-1}]^T$, $\matY = [\vecy_0,\dots,\vecy_{N-1}]^T$, and $\matW = [\vecw_0,\dots,\vecw_{N-1}]^T$. Furthermore, we let $\widehat\matX = \matX\matF_N^H$, $\widehat\matY = \matY\matF_N^H$, $\widehat\matW = \matW\matF_N^H$, and $\widehat\matH_k = \sum_{\ell=0}^{L-1} \matH_\ell e^{-jk\frac{2\pi}{N}\ell}$.
A cyclic prefix of length $L-1$ is prepended to the transmit signal at the \ac{BS}. After removing the cyclic prefix and after a \ac{DFT} at the \acp{UE}, the received signal at the~$U$~\acp{UE} and on the $k$th subcarrier can be written as
\begin{IEEEeqnarray}{rCl} \label{eq:received_freq}
	\hat\vecy_k &=& \widehat\matH_k\hat\vecx_k + \hat\vecw_k	
\end{IEEEeqnarray}
for $k = 0,\dots,N-1$. Here, $\hat\vecx_k$, $\hat\vecy_k$, and $\hat\vecw_k$ correspond to the $k$th column of $\widehat\matX$, $\widehat\matY$, and $\widehat\matW$, respectively.

\subsection{Precoding, Quantization, and OFDM Parameters}

We use the disjoint sets $\setI$ and $\setG$, where $\abs{\setI} + \abs{\setG} = N$, to denote the set of subcarriers associated with information symbols (occupied subcarriers) and  zeros (guard subcarriers), respectively.
We let $S = \abs{\setI}$ be the number of occupied subcarriers and define $N/S$ as the oversampling ratio. Let $\vecs_k = [s_{1,k},\dots,s_{U,k}]^T$ denote the symbol vector associated with the $k$th subcarrier ($k = 0, \dots, N-1$). 
We assume that $s_{u,k} \in \setO$ for all $k \in \setI$  and that $s_{u,k} = 0$ for all $k \in \setG$. Here, $\setO$ represents a \ac{QAM}~constellation (e.g., 16-QAM),.

The precoder uses the available transmit-side channel-state information to map the symbols $\matS = [\vecs_0, \dots, \vecs_{N-1}] \in \setO^{U \times N}$, to the transmitted signal $\matX$, which must satisfy the average power constraint $\Ex{\matS}{\opnorm{\matX}_F^2} = S$.
Due to the finite resolution of real-world \acp{DAC} we require that $\matX \in \setX_{p}^{B \times N}$,  where $\setX_{p}$ is the set of values that are supported by the \acp{DAC}. 
%
%
We shall assume that $\setX_{p}$ is a constant-envelope alphabet and let $p > 0$ be the number of \emph{phase bits}, so that $2^{p}$ is the number of possible phases of the signal transmitted at each antenna. 
Furthermore, we let $|x|^2 ={P_\text{ant}}$, $x\in\setX_{p}$. Here, $P_\text{ant} = S/(BN)$ is the per-antenna transmit power, which ensures that the average power constraint is satisfied.
For $p < \infty$, the $m$th element ($m = 0, \dots, 2^p-1$) of the set $\setX_{p}$ is hence given by $(P_\text{ant})^{1/2} e^{j(\pi + 2\pi m )/2^{p}}$. We let~$\setX_{\infty} =  \Large\{ x \in \opC: \abs{x}^2 = P_\text{ant} \Large\}$.


In this paper, we shall benchmark the performance of our nonlinear precoding algorithm, SQUID-OFDM, against the linear \ac{WF} precoder~\cite{joham05a} given by
\begin{IEEEeqnarray}{rCl} \label{eq:WF_precoding}
	\matX^\text{WF} &=& \setP_{p}\lefto( \widehat\matZ^\text{WF}\matF_N^H\right)\!,
\end{IEEEeqnarray}
where the $k$th column of $\widehat\matZ^\text{WF}$ is given by
\begin{IEEEeqnarray}{rCl} \label{eq:WF_persub}
	\hat\vecz_k^\text{WF} &=& \frac{1}{\beta^\text{WF}}\widehat\matH_k^H\lefto( \widehat\matH_k\widehat\matH_k^H + UN_0\matI_U\right)^{-1}	\! \vecs_k
\end{IEEEeqnarray}
for $k \in \setI$, and by $\hat\vecz_k^\text{WF} = \veczero_{B \times 1}$ for $k \in \setG$.
Here, the constant $\beta^\text{WF} \in \opR^+$ ensures that $\opE_{\matS}\Large[\opnorm{\widehat\matZ^\text{WF}\matF_N^H}_F^2\Large] = S$.
Prior~to~transmission, the time-domain precoded signal $\widehat\matZ^\text{WF}\matF_N^H$~in \eqref{eq:WF_precoding} is quantized by the function $\setP_{p}(\cdot): \opC^{B \times N} \rightarrow \setX_{p}^{B \times N}$, which is applied entry-wise to the matrix $\matX^\text{WF}$, so that the transmitted signal matches the transcoder in the \acp{DAC}.
Specifically,
\begin{IEEEeqnarray}{rCl}
\setP_{p}(z) &=& 	
\begin{cases}
	\sqrt{P_\text{ant}}e^{ j \frac{2\pi}{2^{p}}\lefto(\lefto\lfloor \frac{2^{p}\!\arg(z)}{2\pi}  \right\rfloor + \frac{1}{2}\right)}, & p < \infty\\
	\sqrt{P_\text{ant}} e^{ j \arg(z)}, & p = \infty.
\end{cases} \IEEEeqnarraynumspace \label{eq:quantizer}
\end{IEEEeqnarray}
Note that for the $2$-phase-bit case ($p = 2$), we retrieve from~\eqref{eq:quantizer} the 1-bit-DAC setup studied in~\cite{jacobsson17c}. There, the in-phase and quadrature components of the per-antenna transmitted signal are generated independently by a pair of 1-bit-\ac{DAC} and
\begin{IEEEeqnarray}{rCl}
\setP_{2}(z) =\sqrt{\frac{P_\text{ant}}{2}}\lefto(\sign\lefto(\Re\{z\}\right) + j\sign\lefto(\Im\{z\}\right)\right).	
\end{IEEEeqnarray}
The $1$-phase-bit case ($p=1$), on the other hand, corresponds to the case when there is only \emph{a single} 1-bit DAC per antenna, i.e., the transmitted signal has no in-phase component.

\section{Nonlinear Constant-Envelope Precoding} \label{sec:nonlinear}

As in~\cite{jacobsson17d, jacobsson16d}, we focus on a nonlinear precoding strategy that minimizes the \ac{MSE} at the \acp{UE}. Let $\textit{MSE}_{u,k} = \Ex{w_{u,k}}{\lvert s_{u,k} - \beta \hat{y}_{u,k}\rvert^2}$ denote the \ac{MSE} for the $u$th UE and on the $k$th subcarrier.
Here, $\hat{y}_{u,k}$ is the $u$th element of~$\hat\vecy_k$ and $\beta \in \opR^+$ is a constant that takes into account the channel gain.
With these definitions, we write the sum-\ac{MSE} over the $U$ \acp{UE} and over the $S$ occupied subcarriers as
\begin{IEEEeqnarray}{rCl}
\sum_{u=1}^U\sum_{k \in \setI} \textit{MSE}_{u,k} 
&=& \sum_{k \in \setI} \Ex{\hat\vecw_k}{\opnorm{\vecs_k - \beta\hat\vecy_k}_2^2} \\
&=& \sum_{k \in \setI}\opnorm{\vecs_k - \beta\widehat\matH_k\hat\vecx_k}_2^2 + \beta^2 U S N_0.
\end{IEEEeqnarray}
Recall that $\hat\vecx_k$ is the $k$th column of $\widehat\matX$. We can now define the sum-\ac{MSE}-optimal precoding problem~(PP) as follows:
\begin{IEEEeqnarray}{rCl} \label{eq:problem_wideband}
\text{(PP)} & \quad & 
\left\{\begin{array}{cll}
\!\!\!\underset{\matX \in \setX_{p}^{B \times N}\!,\, \beta \in \reals^+}{\text{minimize}} & \!\!\! \sum\limits_{k \in \setI}\opnorm{\vecs_k - \beta\widehat\matH_k\hat\vecx_k}_2^2 + \beta^2 U S N_0 \\
\!\!\!\text{subject to} & \!\!\! \matX =  \widehat\matX \matF_N^H.
\end{array}\right. \IEEEeqnarraynumspace
\end{IEEEeqnarray}
For constant-envelope signals that adhere to the average power constraint, it holds that $\opnorm{\text{vec}(\matX)}_\infty^2 = P_\text{ant}$, and the problem~(PP) can equivalently be written~as 
\begin{IEEEeqnarray}{rCl} \label{eq:problem_wideband_ce}
&&\begin{array}{cll}
\!\!\!\underset{\matX \in \setX_{p}^{B \times N}\!,\, \beta \in \reals^+}{\text{minimize}} & \sum\limits_{k \in \setI}\opnorm{\vecs_k - \beta\widehat\matH_k\hat\vecx_k}_2^2 + \beta^2 \gamma \opnorm{\text{vec}\lefto(\matX\right)}_\infty^2 \\
\!\!\!\text{subject to} & \matX =  \widehat\matX \matF_N^H,
\end{array} \IEEEeqnarraynumspace
\end{IEEEeqnarray}
where $\gamma = B U N N_0$. Proceeding analogously to~\cite[Sec.~IV-B]{jacobsson17d}, by setting $\widehat\matB = \beta\widehat\matX$ and by dropping the nonconvex constraint $\matX \in \setX_{p}^{B \times N}$, we obtain the following convex relaxation of the problem in~\eqref{eq:problem_wideband_ce}, which we denote by $(\text{P}\ell_{\infty}^2)$:
\begin{IEEEeqnarray}{rCl} \label{eq:problem_inf}
(\text{P}\ell_{\infty}^2) & \,\,\, & 
\underset{\widehat\matB \in \opC^{B \times N}}{\text{minimize}}  \sum\limits_{k \in \setI}\opnorm{\vecs_k \! - \widehat\matH_k\hat\vecb_k}_2^2 + \gamma \opnorm{\text{vec}\big(\widehat\matB\matF_N^H\big) }_{\infty}^2. \ \IEEEeqnarraynumspace
\end{IEEEeqnarray}
Here, $\hat\vecb_k$ is the $k$th column of $\widehat\matB$. 
Let $\widehat\matB^{\text{P}\ell_{\infty}^2}$ and $\beta^{\text{P}\ell_{\infty}^2}$ denote the optimal solutions to the problem ($\text{P}\ell_{\infty}^2$). We obtain the desired matrix $\matX^{\text{P}\ell_{\infty}^2}$ by converting $\widehat\matB^{\text{P}\ell_{\infty}^2}$ to time-domain and by mapping the resulting matrix to the set $\setX_{p}^{B \times N}$ using~\eqref{eq:quantizer},~i.e.,
\begin{IEEEeqnarray}{rCl} \label{eq:quantized_output}
	\matX^{\text{P}\ell_{\infty}^2} &=& \setP_{p}\big( \widehat\matB^{\text{P}\ell_{\infty}^2}\matF_N^H \big).
\end{IEEEeqnarray}
In~\fref{sec:SQUID}, we will show that the problem ($\text{P}\ell_{\infty}^2$) can be solved efficiently. 
Note that for the 2-phase-bit case, it holds that $\opnorm{\text{vec}\lefto( \matX \right)}_\infty^2 = 2\opnorm{\text{vec}\lefto( \matX \right)}_{\widetilde\infty}^2$. In this case, it turns out that one achieves better performance by solving instead the following optimization~problem, which we denote by $(\text{P}\ell_{\widetilde\infty}^2)$:
\begin{IEEEeqnarray}{rCl} \label{eq:problem_inftilde}
(\text{P}\ell_{\widetilde\infty}^2) & \ & 
\underset{\widehat\matB \in \opC^{B \times N}}{\text{minimize}}  \sum\limits_{k \in \setI}\opnorm{\vecs_k \! - \widehat\matH_k\hat\vecb_k}_2^2 + 2 \gamma \opnorm{\text{vec}\big(\widehat\matB\matF_N^H\big) }_{\widetilde\infty}^2. \ \IEEEeqnarraynumspace
\end{IEEEeqnarray}
We shall discuss the implications of this slight modification of the precoding problem in the next section.

\subsection{SQUID-OFDM Precoding} \label{sec:SQUID}

Douglas-Rachford splitting~\cite{lions79a} is an efficient iterative scheme to solve convex optimization problems of the form
\begin{align} \label{eq:problem_rewrite}
\begin{array}{lll}
	\underset{\widehat\matB \in \opC^{B \times N}}{\text{minimize}} & f\big(\widehat\matB\big) + g\big(\widehat\matB\big),
\end{array}
\end{align}
where $f(\cdot)$ and $g(\cdot)$ are closed convex functions, which have \emph{proximal operators}~\cite{parikh13a} defined as follows:
\begin{IEEEeqnarray}{rCl}
\text{prox}_{f}(\matV) &=& \argmin_{\widehat\matB \in \opC^{B \times N}} f(\widehat\matB) +\textstyle  \frac{1}{2}\opnorm{\widehat\matB - \matV}_F^2 \label{eq:prox_f} \\
\text{prox}_{g}(\matV) &=& \argmin_{\widehat\matB \in \opC^{B \times N}} g(\widehat\matB) + \textstyle \frac{1}{2}\opnorm{\widehat\matB - \matV}_F^2. \label{eq:prox_g} 
\end{IEEEeqnarray}
By starting at an arbitrary $\widehat\matB^{(0)}$ and $\widehat\matC^{(0)}$, Douglas-Rachford splitting solves problems of the form~\eqref{eq:problem_rewrite} {exactly}~\cite{eckstein92a} by repeating for $t = 1, \dots, T$, where $T$ is the maximum number of iterations,  the following iterative procedure:
\begin{IEEEeqnarray}{rCl}
\widehat\matA^{(t)} &=& \text{prox}_f \lefto( 2\widehat\matB^{(t-1)} - \widehat\matC^{(t-1)}\right) \label{eq:DR_first} \\
\widehat\matB^{(t)} &=& \text{prox}_g\lefto( \widehat\matC^{(t-1)} + \widehat\matA^{(t)} - \widehat\matB^{(t-1)}\right) \\
\widehat\matC^{(t)} &=& \widehat\matC^{(t-1)} + \widehat\matA^{(t)}  - \widehat\matB^{(t)}.
\end{IEEEeqnarray}

We now outline the {SQUID-OFDM} precoder, which builds upon the SQUID precoder proposed in~\cite[Sec.~IV-B]{jacobsson17d} and performs Douglas-Rachford splitting to solve the problems ($\text{P}\ell_{\infty}^2$) and  ($\text{P}\ell_{\widetilde\infty}^2$). Specifically, SQUID-OFDM extends \ac{SQUID} to support OFDM, oversampling DACs, and arbitrary constant-envelope alphabets.
Let $f\big(\widehat\matB\big) = \sum_{k \in \setI} \opnorm{\vecs_k - \widehat\matH_k\hat\vecb_k}_2^2$. For the problem~($\text{P}\ell_{\infty}^2$), $g\big(\widehat\matB\big) = \gamma \opnorm{\text{vec}\big(\widehat\matB\matF_N^H\big)}_{\infty}^2$. For the problem~($\text{P}\ell_{\widetilde\infty}^2$), $g\big(\widehat\matB\big) = 2 \gamma \opnorm{\text{vec}\big(\widehat\matB\matF_N^H\big)}_{\widetilde\infty}^2$. 
In both cases, the proximal operator for $g(\cdot)$ in~\eqref{eq:prox_g} can computed using~\cite[Alg.~1]{jacobsson17d}. For the proximal operator of $f(\cdot)$, we note that the objective function in~\eqref{eq:prox_f} is separable in the columns of $\widehat\matB$ and that the $k$th column of $\widehat\matA^{(t)}$ in \eqref{eq:DR_first} can be~computed~as
\begin{IEEEeqnarray}{rCl}
	\veca_k^{(t)} &=& \textstyle \Big( \widehat\matH_k^H\widehat\matH_k \! + \frac{1}{2}\matI_B \! \Big)^{\!-1} \! \Big( \widehat\matH_k^H \vecs_k +  \hat\vecb_k^{(t-1)} \! - \frac{1}{2}\hat\vecc_k^{(t-1)} \Big) \IEEEeqnarraynumspace \\
	&=& \big(\matI_B - \matQ_k \widehat\matH_k \big) \big( 2\hat\vecb_k^{(t-1)} \! - \hat\vecc_k^{(t-1)} \big) + \vecd_k \label{eq:prox_faster}
\end{IEEEeqnarray}
for $k \in \setI$, and $\veca_k^{(t)} = 2\hat\vecb_k^{(t-1)}  - \hat\vecc_k^{(t-1)}$ for $k \in \setG$. Here, $\veca_k^{(t)}$ is the $k$th column of $\widehat\matA^{(t)}$ and $\vecc_k^{(t)}$ is the $k$th column of $\widehat\matC^{(t)}$. To derive~\eqref{eq:prox_faster}, we used the Woodbury matrix identity to reduce the dimension of the inverse and to speed up computations by precomputing, for $k \in \setI$, the matrix $\matQ_k \in \opC^{B \times U}$ and the vector $\vecd_k \in \opC^{B}$, which are defined as~follows:
\begin{IEEEeqnarray}{rCl} 
\matQ_k &=& \widehat\matH_k^H \Big(\widehat\matH_k \widehat\matH_k^H + \frac{1}{2}\matI_U \Big)^{-1} \label{eq:matrixQ}\\
\vecd_k &=& 2 \lefto( \widehat\matH_k^H\vecs_k - \matQ_k\widehat\matH_k\widehat\matH_k^H \vecs_k\right) . \label{eq:vectorv}
\end{IEEEeqnarray}
We can now solve the problems~($\text{P}\ell_{\infty}^2$) and~($\text{P}\ell_{\widetilde\infty}^2$) by using the iterative procedure outlined in~\fref{alg:squid}.
\begin{oframed}
\begin{alg}[SQUID-OFDM]\label{alg:squid}  
Compute $\matQ_k$ and $\vecd_k$ for $k \in \setI$ using \fref{eq:matrixQ} and \fref{eq:vectorv}, respectively. Initialize $\widehat\matB^{(0)} = \veczero_{B \times N}$, and $\widehat\matC^{(0)} = \veczero_{B \times N}$.
Then, at every iteration $t=1,2,\dots,T$, compute the following three quantities:
\begin{IEEEeqnarray}{rCl} \label{eq:ai_expanded}
&&\hat\veca_k^{(t)}	=  
\begin{cases}
\matR_k \Big( 2\hat\vecb_k^{(t-1)}  - \hat\vecc_k^{(t-1)} \Big) + \vecd_k,  & k \in \setI \\
2\hat\vecb_k^{(t-1)}  - \hat\vecc_k^{(t-1)}, & k \in \setG	
\end{cases}
 \IEEEeqnarraynumspace \label{eq:ai} \\
&&\widehat\matB^{(t)} = \text{prox}_{g} \lefto( \widehat\matC^{(t-1)} + \widehat\matA^{(t)}  - \widehat\matB^{(t-1)}\right) \matF_N\label{eq:bi} \IEEEeqnarraynumspace \\
&&\widehat\matC^{(t)} = \widehat\matC^{(t-1)} + \widehat\matA^{(t)} - \widehat\matB^{(t)}. \IEEEeqnarraynumspace 
\end{IEEEeqnarray}
Here, $\matR_k = \big(\matI_B - \matQ_k \widehat\matH_k \big)$. In~\eqref{eq:bi}, $\text{prox}_g(\cdot)$ is computed using~\cite[Alg.~1]{jacobsson17d}.
After the last iteration, obtain the transmitted signal $\matX^{(T)}$ by quantizing $\widehat\matB^{(T)}\matF_N^H$ to the constant-envelope alphabet~$\setX_{p}^{B \times N}$ using~\eqref{eq:quantizer}.
\end{alg}
\end{oframed}

\begin{figure*}[tp]
\centering
\subfloat[1-phase-bit SQUID-OFDM.]{\includegraphics[width = .22\textwidth]{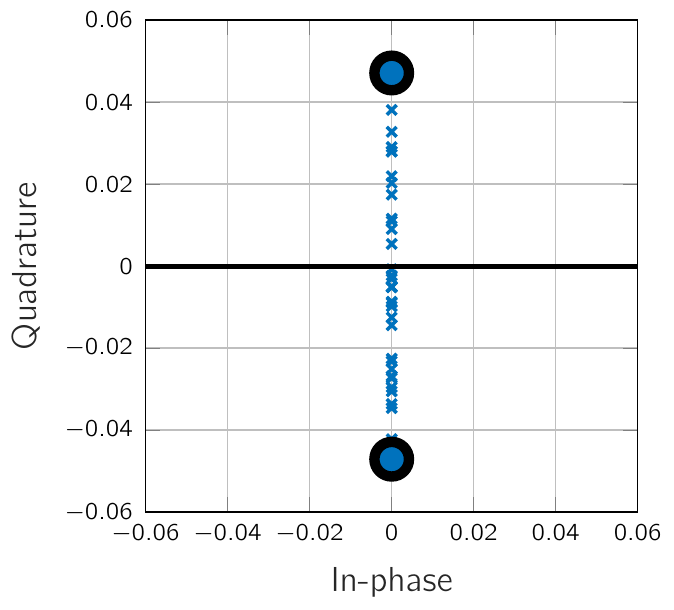}\label{fig:infreal}} \qquad
\subfloat[2-phase-bit SQUID-OFDM.]{\includegraphics[width = .22\textwidth]{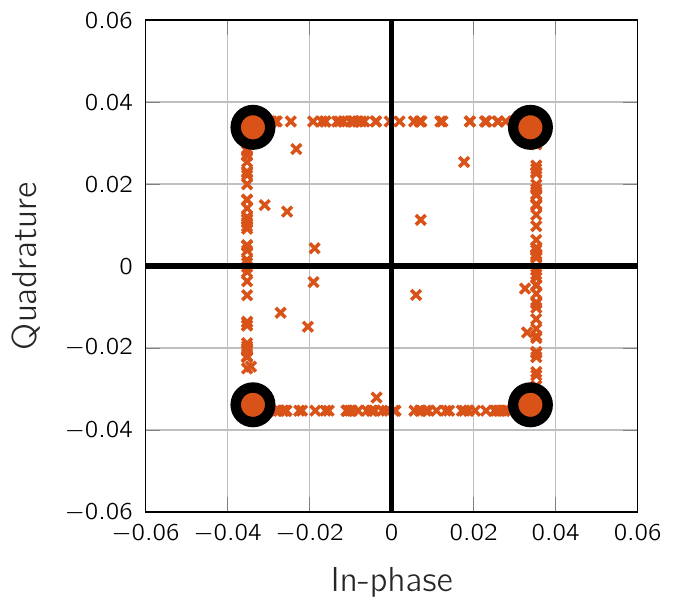}\label{fig:inftilde}} \qquad
\subfloat[3-phase-bit SQUID-OFDM.]{\includegraphics[width = .22\textwidth]{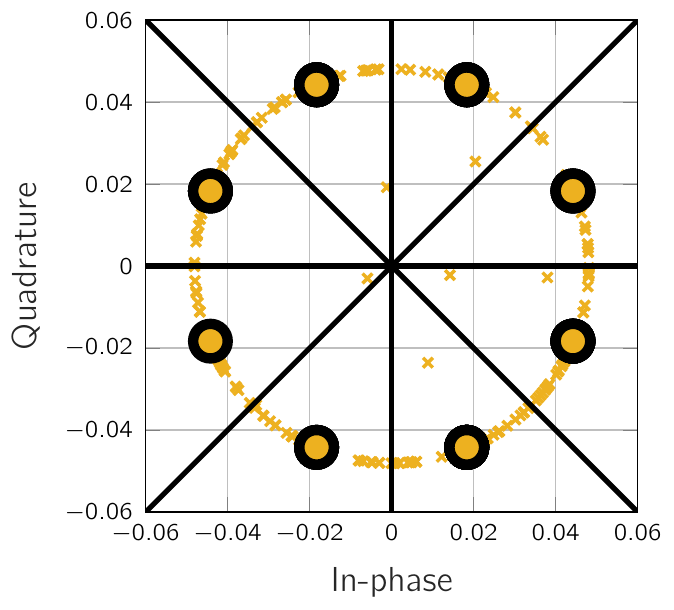}\label{fig:inf}}
\caption{Per-antenna SQUID-OFDM output before and after quantization for 16-QAM, $T = 20$, $\textit{SNR} = 10$\,dB, $B = 128$, $U = 16$, $S = 1200$, and $N = 4096$. The cross-markers correspond to the output before quantization, the  circles to the quantized output, and the lines to the decision regions for the~quantizer.}
\label{fig:transmit_signal}
\vspace{-0.2cm}
\end{figure*}

Fig.~\ref{fig:transmit_signal} shows the time-domain SQUID-OFDM output $\widehat\matB^{(20)}\matF_N^H$  after $T = 20$ iterations before and after quantization.\footnote{The simulation parameters are given in~\fref{sec:sim_param}.}
Fig.~\ref{fig:inftilde} shows the SQUID-OFDM output for the problem  ($\text{P}\ell_{\widetilde\infty}^2$) before and after 2-phase-bit quantization.\footnote{By setting $S=1$ and $N=1$ for the problem ($\text{P}\ell_{\widetilde\infty}^2$), \fref{alg:squid} reduces to the SQUID precoder for single-carrier transmission~\cite[Sec.~IV-B]{jacobsson17d}.} We see that the $\ell_{\widetilde\infty}$-norm constrains the SQUID-OFDM output to a box in the complex plane, which limits the error caused by the quantizer.
Fig.~\ref{fig:inf} shows the output for the problem ($\text{P}\ell_{\infty}^2$) before and after 3-phase-bit quantization. In this case, the $\ell_{\infty}$-norm constrains the SQUID-OFDM output to a circle in the complex plane, which is suitable for quantization using three phase bits or more. 
Fig.~\ref{fig:infreal} shows the SQUID-OFDM output for the problem ($\text{P}\ell_{\infty}^2$) before and after 1-phase-bit quantization. Here, we slightly modified the problem to force the real part of the output to the proximal operator~$\text{prox}_g(\cdot)$ to zero. This constrains the SQUID-OFDM output to a line in the complex plane, which is suitable for quantization using only one phase~bit.

\subsection{Computational Complexity}

\fref{tab:complexity} shows the computational complexity characterized by the number of \emph{real-valued multiplications} for SQUID-OFDM and WF precoding. In what follows, we assume that one complex-valued multiplication requires four real-valued~multiplications. 

\setlength{\textfloatsep}{10pt}
\begin{table}[tp]
\renewcommand{\arraystretch}{1.05}
\centering
\begin{minipage}[c]{1\columnwidth}
\vspace{-0.1cm}
\centering
\caption{Complexity comparison between WF and SQUID-OFDM.}
\label{tab:complexity}
\begin{tabular}{@{}lc@{}}
\toprule
Precoder   &  Computational complexity \\
\midrule
\multirow{2}{*}{WF} & $ 2S\lefto(\frac{1}{3}U^3 + BU^2 + 2U^2 -\frac{1}{3}U\right)$ \\ &  $+ 4B\lefto(N\log_2 N - 3N + 4\right)$ \\
\multirow{2}{*}{SQUID-OFDM} & $2S\lefto(\frac{5}{3}U^3 + 3BU^2 + \lefto(6B-\frac{2}{3}\right)U\right)$ \\ &  $+ 4TB(2SU + 2N\log_2{N} - 5N + 8)$ \\
\bottomrule
\end{tabular}
\end{minipage}
\end{table}

\subsubsection{WF Precoding} Computing~\fref{eq:WF_persub} exactly for all $k \in \setI$ using implicit Cholesky-based matrix inversion~\cite{wu17a} requires $2S\lefto(\frac{1}{3}U^3 + BU^2 + 2U^2 -\frac{1}{3}U\right)$ real-valued multiplications. At each antenna element, the frequency-domain precoded vector is converted to time-domain via an \ac{IDFT}. Computing these \acp{IDFT} require $4B\lefto(N\log_2 N - 3N + 4\right)$ real-valued multiplications if the \acp{IDFT} are computed using the split-radix \ac{FFT} algorithm~\cite[Sec.~4.3]{duhamel90a}. By adding these two numbers, we obtain the complexity reported in~\fref{tab:complexity}.

\subsubsection{SQUID-OFDM} The preprocessing step of SQUID-OFDM involves computing~\eqref{eq:matrixQ} and~\eqref{eq:vectorv} for $k \in \setI$.
Proceeding as in~\cite{li17d}, we find that computing $\matQ_k$ in~\eqref{eq:matrixQ} for $k \in \setI$ requires $2S(\frac{5}{3}U^3 + 3BU^2 - \frac{2}{3}U)$ real-valued multiplications. 
Furthermore, computing $\vecd_k$ in~\eqref{eq:vectorv} for $k \in \setI$  requires an additional $12SBU$ real-valued multiplications.
By adding these numbers, we find that the preprocessing complexity of \ac{SQUID}-OFDM is $2S\lefto(\frac{5}{3}U^3 + 3BU^2 + \lefto(6B-\frac{2}{3}\right)U\right)$. 
Moving on to the per-iteration complexity. Computing efficiently the vectors $\veca_k^{(t)}$ in~\eqref{eq:ai_expanded} for $k \in \setI$  requires $8SBU$ real-valued multiplications per iteration.\footnote{Efficiently computing the step~\eqref{eq:ai_expanded} for $k \in \setI$ involves first computing $\vecv_k = 2\hat\vecb_k^{(t-1)} - \hat\vecc_k^{(t-1)}$ and then computing $\hat\veca_k^{(t)} = \vecv_k - \matQ_k \widehat\matH_k\vecv_k + \vecd_k$.}
Furthermore, executing \cite[Alg.~1]{jacobsson17d} requires $4BN$ real-valued multiplications, which means that computing $\widehat\matB^{(t)}$ in \eqref{eq:bi}, if the split-radix \ac{FFT} algorithm is used to compute the \ac{IDFT} and \ac{DFT}, requires an additional $4B(2N\log_2{N} - 5N + 8)$ real-valued multiplications per iteration.
Hence, the per-iteration complexity of SQUID-OFDM is $4B(2SU + 2N\log_2{N} - 5N + 8)$.\footnote{For the single-carrier case (i.e., when $N = 1$ and $S=1$), SQUID-OFDM reduces to single-carrier SQUID~\cite[Sec.~IV-B]{jacobsson17d} and the per-iteration complexity reduces to $8BU + 4B$ real-valued multiplications (no split-radix \ac{FFT} algorithm has to be computed for this case).}  
Finally, by adding the preprocessing complexity and the per-iteration complexity, we obtain the complexity for $T$ iterations reported in~\fref{tab:complexity}. Note that the computational complexity of both SQUID-OFDM and WF precoding scales \emph{linearly} in the number of BS antennas~$B$.

\section{Numerical Results} \label{sec:numerical}

%

\subsection{Simulation Parameters} \label{sec:sim_param}

Due to space constraints, we focus on a selected set of system parameters.\footnote{Our simulation framework is available for download from GitHub (https://github.com/quantizedmassivemimo/1bit\_precoding\_ofdm).}
Specifically, we consider the case in which the \ac{BS} is equipped with $B = 128$ antennas and serves $U =16$ \acp{UE}. We consider \ac{LTE}-inspired \ac{OFDM} parameters~\cite{dahlman13a} with $S = 1200$ occupied subcarriers and where $N = 4096$ (the oversampling ratio is $N/S = 4096/1200 \approx 3.41$). The subcarrier spacing is $\Delta f = 15$~kHz and the sampling rate is $f_s = N\Delta f = 61.44$~MHz.
The set of occupied subcarriers is $\setI = \{1, 2, \dots, 600,  3497, 3498,  \dots, 4096\}$ and the set of guard subcarriers is $\setG = \{0,1, \dots, 4096 \} \!\setminus \setI$. 
The entries of~$\{ \matH_\ell \}_{\ell = 0}^{L-1}$ are i.i.d.\ $\jpg\big(0, 1/L \big)$ (Rayleigh fading). The number of taps is $L = 4$.
We furthermore assume that~the $u$th \ac{UE} ($u = 1, 2, \dots, U$) scales the received signal for each \ac{OFDM} symbol~by~\cite{jacobsson16d}
\begin{IEEEeqnarray}{rCl} \label{eq:beta_u}
\beta_u = \frac{1}{\sqrt{\frac{1}{S}\sum_{k \in \setI} \abs{\hat{y}_k}^2 - N_0}},
\end{IEEEeqnarray}
to obtain an estimate $\tilde{s}_{u,k} = \beta_u \hat{y}_{u,k}$ of $s_{u,k}$ for $k \in \setI$.

\subsection{Convergence and Complexity}

\begin{figure}
\centering
\includegraphics[width = .85\columnwidth]{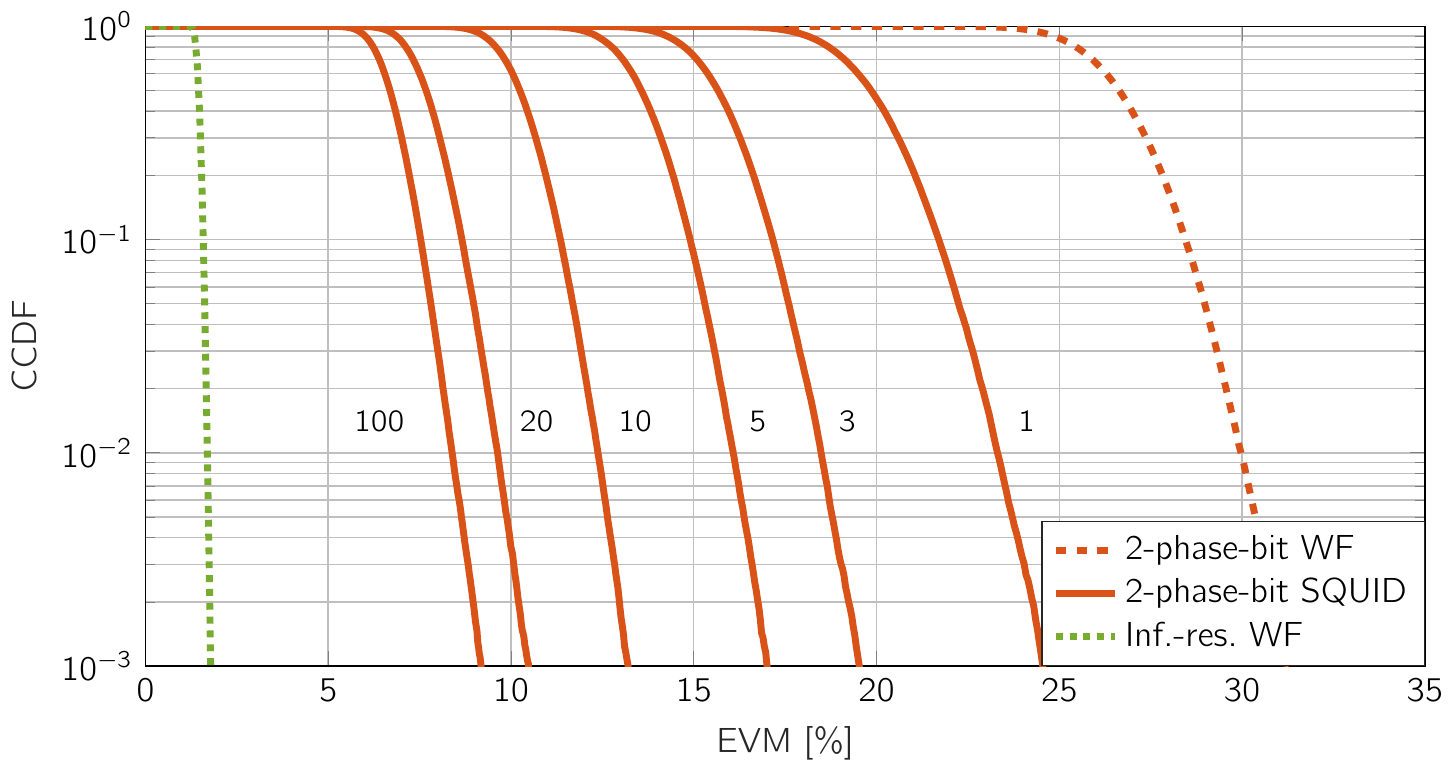}	
\caption{CCDF of the EVM with 16-QAM; $\textit{SNR} = 10$\,dB, $B = 128$, $U = 16$, $S = 1200$, and $N = 4096$.  The number next to the CCDF curves corresponds to the number of iterations for \ac{SQUID}-OFDM.}
\label{fig:evm}
\end{figure}

We start by investigating the convergence of \ac{SQUID}-OFDM for the 2-phase-bit case. Fig.~\ref{fig:evm} shows the \ac{CCDF} of the \ac{EVM}, with 16-QAM signaling for $\textit{SNR} = 10$\,dB. Here, the \ac{EVM} for the $u$th \ac{UE} ($u = 1,\dots,U$)~is defined as
\begin{IEEEeqnarray}{rCl}
	\textit{EVM}_u &=& \sqrt{\frac{\sum_{k \in \setI}\big\lvert s_{u,k} - \beta_u \hat\vech_{u,k}^T \hat\vecx_k \big\rvert^2}{\sum_{k \in \setI}\abs{s_{u,k}}^2}},
\end{IEEEeqnarray}
where $\hat\vech_{u,k}^T$ is the $u$th row of $\widehat\matH$ and where $\beta_u$ is given by~\eqref{eq:beta_u}.
For reference, we also show the \ac{CCDF} of the \ac{EVM} with \ac{WF} precoding for the 2-phase-bit case and for the \emph{infinite-resolution} case (i.e., when $\matX^\text{WF} = \widehat\matZ^\text{WF}\matF_N^H \in \opC^{B \times N}$), respectively. Interestingly, we see that SQUID-OFDM with only one iteration already significantly outperforms WF precoding in terms of \ac{EVM}. Furthermore, we see from \fref{tab:complexity} that for $T=1$, the complexity of SQUID-OFDM is just about  $3$ times higher than that of WF precoding. 
We also see from Fig.~\ref{fig:evm} that by increasing the number of iterations from $20$ to $100$, SQUID-OFDM attains only marginal \ac{EVM} gains. In what follows, we set the number of iterations to $T = 20$. In this case, \ac{SQUID}-OFDM requires approximately $14$ times more real-valued multiplications than the \ac{WF} precoder, assuming the parameters given in~\fref{sec:sim_param}.

\setlength{\textfloatsep}{10pt}

\subsection{Error-Rate Performance}

\setlength{\textfloatsep}{10pt}
\begin{figure}[t]
\centering
\subfloat[Uncoded BER with 4-QAM. SQUID-OFDM outperforms WF precoding irrespectively of the SNR and the number of phase bits.]{\includegraphics[width = .85\columnwidth]{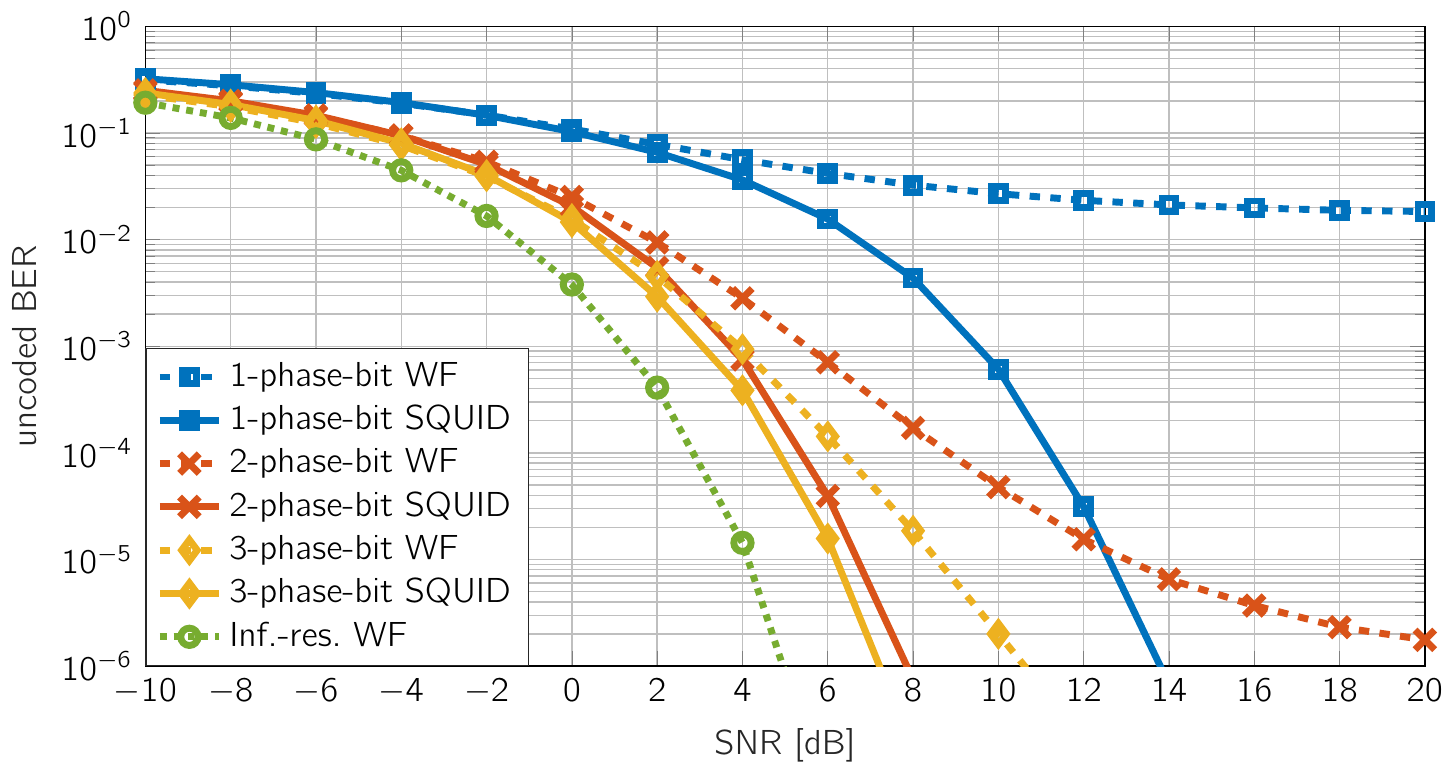}\label{fig:ber_qpsk}} \\
\subfloat[Coded BER with 16-QAM (rate-5/6 convolutional code). Low coded BERs are supported with SQUID-OFDM.]{\includegraphics[width = .85\columnwidth]{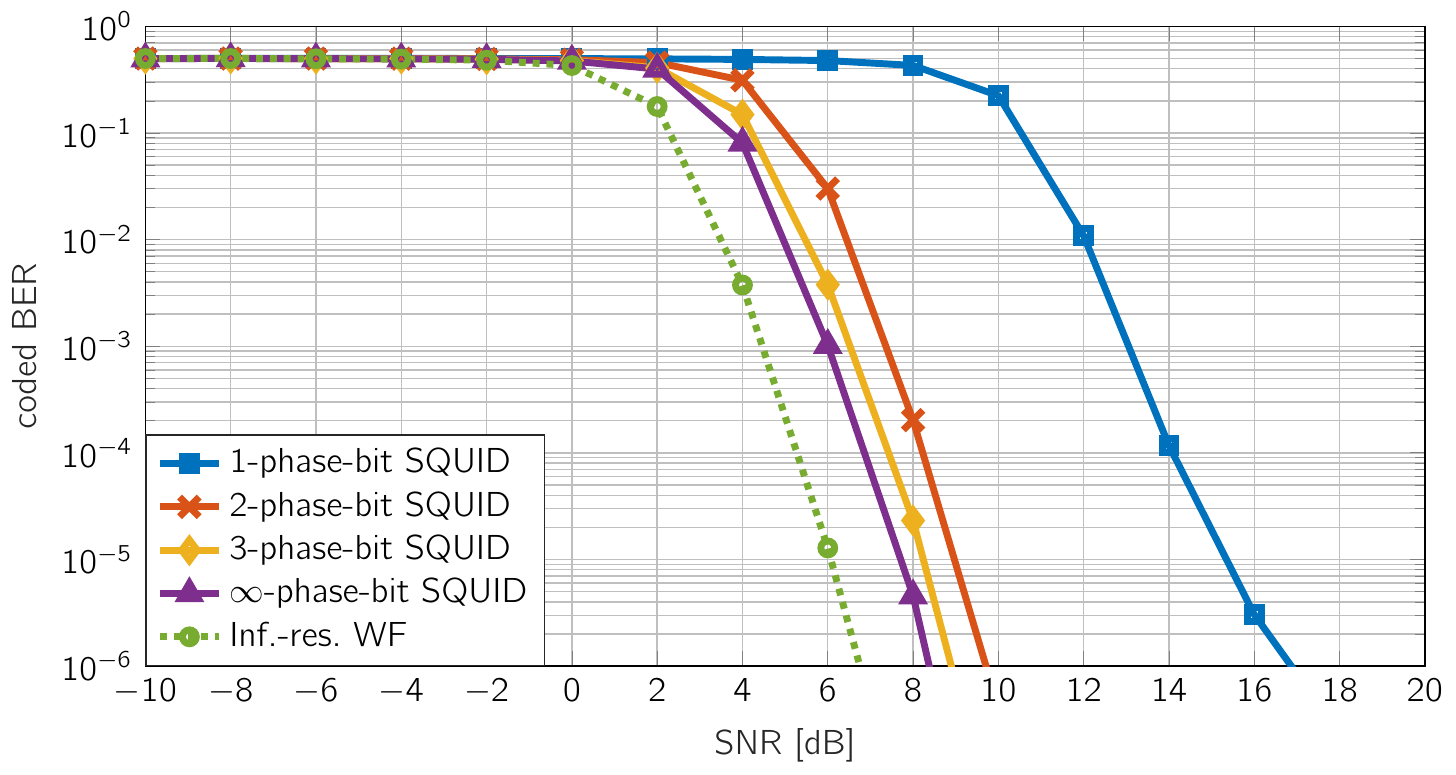}	\label{fig:ber_16qam}}
\caption{Uncoded/coded BER as a function of SNR and the number of phase bits; $B = 128$, $U=16$, $S = 1200$, and $N = 4096$. SQUID-OFDM significantly outperforms linear precoders and approaches infinite resolution~performance.}
\end{figure}

\subsubsection{Uncoded BER} Fig.~\ref{fig:ber_qpsk} shows the uncoded BER with  4-\ac{QAM} for $p$-phase-bit ($p\in\{1,2,3\}$) SQUID-OFDM and WF precoding as a function of the SNR. We also show the uncoded BER with infinite-resolution WF precoding. We assume that the \acp{UE} perform symbol-wise nearest-neighbor decoding (i.e., each UE maps the received signal to the nearest constellation point in~$\setO$). We note that SQUID-OFDM outperforms WF precoding for all considered values of SNR and irrespectively of the number of phase bits. 
Interestingly, low uncoded \acp{BER} are supported even by 1-phase-bit~SQUID-OFDM.

\subsubsection{Coded BER} Fig.~\ref{fig:ber_16qam} shows the coded BER with 16-QAM for $p$-phase-bit ($p\in\{1,2,3, \infty\}$) SQUID-OFDM as a function of the SNR. For $\infty$-phase-bit \ac{SQUID}-OFDM, the output after the last iteration is mapped to the set $\setX_\infty^{B \times N}$ (no quantization). 
At the BS, the information bits are encoded using a weak rate-5/6 convolutional code. Each codeword is randomly interleaved over $4800$ bits (i.e., over the $S = 1200$ occupied subcarriers in an OFDM symbol). To detect the information bits, each \ac{UE} performs soft-input max-log~BCJR decoding. We note that low coded BERs are supported with SQUID-OFDM. Also, we note that $2$-phase-bit SQUID-OFDM already offers performance close to that of~$\infty$-phase-bit~SQUID-OFDM.

\section{Conclusions} \label{sec:conclusions}

We have proposed a nonlinear phase-quantized precoder called SQUID-OFDM for the massive MU-MIMO-OFDM downlink. The precoder extends the \ac{SQUID} precoder in~\cite{jacobsson17d} to support OFDM, oversampling DACs, and arbitrary constant-envelope alphabets.
SQUID-OFDM is shown to offer superior error-rate performance to linear precoders such as \ac{WF} precoding at an increased computational complexity (three times or higher depending on the number of algorithm iterations).
The constant-envelope transmit signals generated by SQUID-OFDM enable energy-efficient \acp{PA}, which is in stark contrast to the infinite-precision \ac{WF} precoder whose \ac{PAR} approximately ranges between $10$\,dB and $12$\,dB. 
Furthermore, for 2-phase-bit SQUID-OFDM, the amount of raw data that has to be fed to the \acp{DAC} is $15.7$\,Gbit/s.
In contrast, a traditional system that uses high-resolution \acp{DAC}, e.g., $12$-bit, must sustain raw baseband data rates that exceed $188$\,Gbit/s for the parameters considered in this~paper, which seems impractical.


\bibliographystyle{IEEEtran}
\begin{spacing}{.97}
\bibliography{IEEEabrv,confs-jrnls,publishers,svenbib}	
\end{spacing}

\end{document}